\documentclass[12pt]{spieman}  
\usepackage{amsmath,amsfonts,amssymb}
\usepackage{graphicx}
\usepackage{setspace}
\usepackage{tocloft}

\usepackage{comment}
\usepackage{soul}

\title{Design of a frequency-independent optic axis Pancharatnam-based achromatic half-wave plate}

\author[a]{Kunimoto Komatsu}
\author[a]{Hirokazu Ishino}
\author[b]{Nobuhiko Katayama}
\author[b]{Tomotake Matsumura}
\author[b]{Yuki Sakurai}
\affil[a]{Department of Physics, Okayama University, Okayama, Japan}
\affil[b]{Kavli Institute for Physics, Mathematics for Universe (WPI), the University of Tokyo, Chiba, Japan}

\cftpagenumbersoff{figure}
\cftpagenumbersoff{table} 
\begin{document} 
\maketitle

\begin{abstract}
Pancharatnam-based achromatic half-wave plates (AHWP) achieve high polarization efficiency over a broad waveband.
These AWHPs generally contain a property whereby the optic axis is dependent on the electromagnetic frequency of the incident radiation.
When the AHWP is used to measure incident polarized radiation with a finite detection bandwidth, this frequency dependence causes an uncertainty in the determination of the polarization angle due to the limited knowledge of the shape of the source spectrum and detection band.
To mitigate this problem, we propose new designs of the AHWP which eliminate the frequency dependence of the optic axis over the bandwidth whilst maintaining high modulation efficiency. 
We carried out this optimization by tuning the relative angles among the individual half-wave plates of the five and nine layer AHWPs. 
The optimized set of relative angles achieves a frequency-independent optic axis over the fractional bandwidth, a bandwidth over which polarization efficiency is greater than 0.9, of 1.3 and 1.5 for the five and nine layer AHWPs, respectively. 
We also study the susceptibility of the alignment accuracy on the polarization efficiency and the frequency dependence of the optic axis, which provides a design guidance for each application. 
\end{abstract}


{\noindent \footnotesize\textbf{*}Kunimoto Komatsu,  \linkable{k.komatsu@s.okayama-u.ac.jp} }

\begin{spacing}{1}   

\section{Introduction}

\label{sect:intro}  
The measurement of the cosmic microwave background (CMB) has played an important role in modern cosmology. 
In recent years, the required sensitivity of CMB polarization telescopes have been raised significantly in order to probe nano-Kelvin fluctuations on top of the CMB temperature, $2.72548 \pm 0.00057$K~\cite{Fixsen_2009}.
This sensitivity allows for the measurement of the tensor-to-scalar ratio of $r>10^{-3}$, or to improve the current upper limit of 
$r<0.044$ obtained from Planck with BICEP2/Keck 2015 data~\cite{tristram2020planck}.
One of the key requirements for a CMB telescope is to be able to probe large angular scales, $\ell<10$, where the inflationary signal at the level of $r\sim 10^{-3}$ is more dominant than that of gravitational lensing.
Correspondingly, instrumental stability is required over long timescales during its scan, and suppression of systematic effects at these large angular scales is essential. 
A continuously-rotating half-wave plate (HWP) can relax some of these stringent requirements.
It is or will be implemented in MAXIPOL~\cite{maxipol}, ABS~\cite{abs}, EBEX~\cite{ebex}, SPIDER~\cite{spider}, POLARBEAR~\cite{polarbear}, POLARBEAR2~\cite{Kaneko_2020,charles_spie}, LSPE/SWIPE~\cite{lspe_swipe}, SO~\cite{so}, and LiteBIRD~\cite{hazumi_spie2020}.
A continuously-rotating HWP in an optical path of the telescope modulates the incident polarized signal by rotating its polarization angle, enabling the use of lock-in techniques to the incident polarized signal.
As a result, the signal band of the linearly polarized light can be up-converted above the timescale of the system instability~\cite{abs}.
Also, the requirement of avoiding any mismatch of the detector properties between detector pairs, e.g. beam, gain, and bandpass, can be relaxed because an individual detector measures the full linear polarization by polarization angle rotation~\cite{I2P_essinger}. 
While a HWP can play a crucial role in mitigating the systematic effects in CMB polarimetry, the HWP itself must also adhere to stringent requirements.

There is no area of sky which is free from Galactic foreground emissions.
Therefore, it is essential to observe over a broad band and also to subtract contributions from foreground emissions by using the difference of spectral shapes. 
A HWP introduces a phase difference of $\pi$ radians between orthogonal linear polarizations.
A common and simple way of producing a HWP is to use a birefringent material. 
In this way, the phase difference can only be exactly $\pi$ at a single frequency determined by the thickness and refractive index of the birefringent material. 
At other frequencies, a conversion efficiency from linear-to-linear polarization states is reduced, which leads to the degradation of polarized sensitivity of a polarimeter.
On the other hand, recent CMB polarization experiments require a fractional bandwidth of a HWP of above 1.0 for ground telescopes and above 1.3 for balloon and space missions.~\cite{charles_spie,so,hazumi_spie2020} 
A conventional solution to increase the bandwidth is to use a recipe proposed by Pancharatnam, which employs three HWPs stacked together with specific relative angles between them~\cite{pancharatnam_1,pancharatnam_2}.
In this paper, we call the multi-layered HWP that follows the Pancharatnam recipe an achromatic HWP (AHWP).
Further investigations have addressed the extension of this original work to the five and nine layered AHWP~\cite{savini,tmatsumura,kkomatsu_jatis_2019}. 
From a calculation, a single layer HWP can achieve about 0.4 of the fractional bandwidth for polarization efficiency of above 0.9, and an AHWP covers the fractional bandwidth of 1.2, 1.3, 1.4, and 1.6 for 3, 5, 7, and 9 layers, respectively~\cite{hanany,savini,kkomatsu_spie_design_2020,kkomatsu_jatis_2019}.

While the AHWP achieves a broader bandwidth of polarization efficiency, one caveat is that it introduces a frequency dependence on the effective fast axis of the AHWP.
This means that a polarimeter using an AHWP has a frequency-dependent polarization angle sensitivity.
$Q$ and $U$ signals defined in a telescope coordinate is are no longer defined by the physical orientation of a polarization-sensitive axis, and it varies as a function of the observing electromagnetic frequency. 
In principle, such an effect can be corrected with perfect knowledge of a spectral response of the instrument and source. 
Any limit in this knowledge can lead to uncertainty of the polarization angle sensitive orientation. 
Past simulation based studies address the importance of this effect.~\cite{bao,bao2,adriaan}
Bao et al. shows that the frequency-dependent polarization angle rotation effect using a conventional AHWP can be neglected when dust is polarized at a level of $>5$\% and $r<0.05$ in case of EBEX.
Furthermore, Duivenvoorden et al. addresses the systematic effect of the HWP including the effectiveness of the frequency-independent AHWP for a future CMB project.
The analysis-based mitigation methods have also been proposed~\cite{max,clara}.
The typical required accuracy of the polarization angle for future inflationary B-mode CMB polarization experiments probing the tensor-to-scalar ratio of $10^{-3}$ is in the range of $1 - 10$~arcmin for both absolute and relative angles~\cite{ysakurai_spie2020, fabio_spie2020, charles_spie, so2}
This is a stringent calibration requirement, and it is desirable to not have to take into account the additional effect from the frequency dependent optic axis. 
In this paper, we have studies the numerical optimization of the AHWP design to eliminate the spectral dependence of the effective fast optic axis of an AHWP. 

\section{HWP Polarimetry}
\subsection{Formalism}
\begin{figure}[t]
\begin{center}
\includegraphics[width=\hsize]{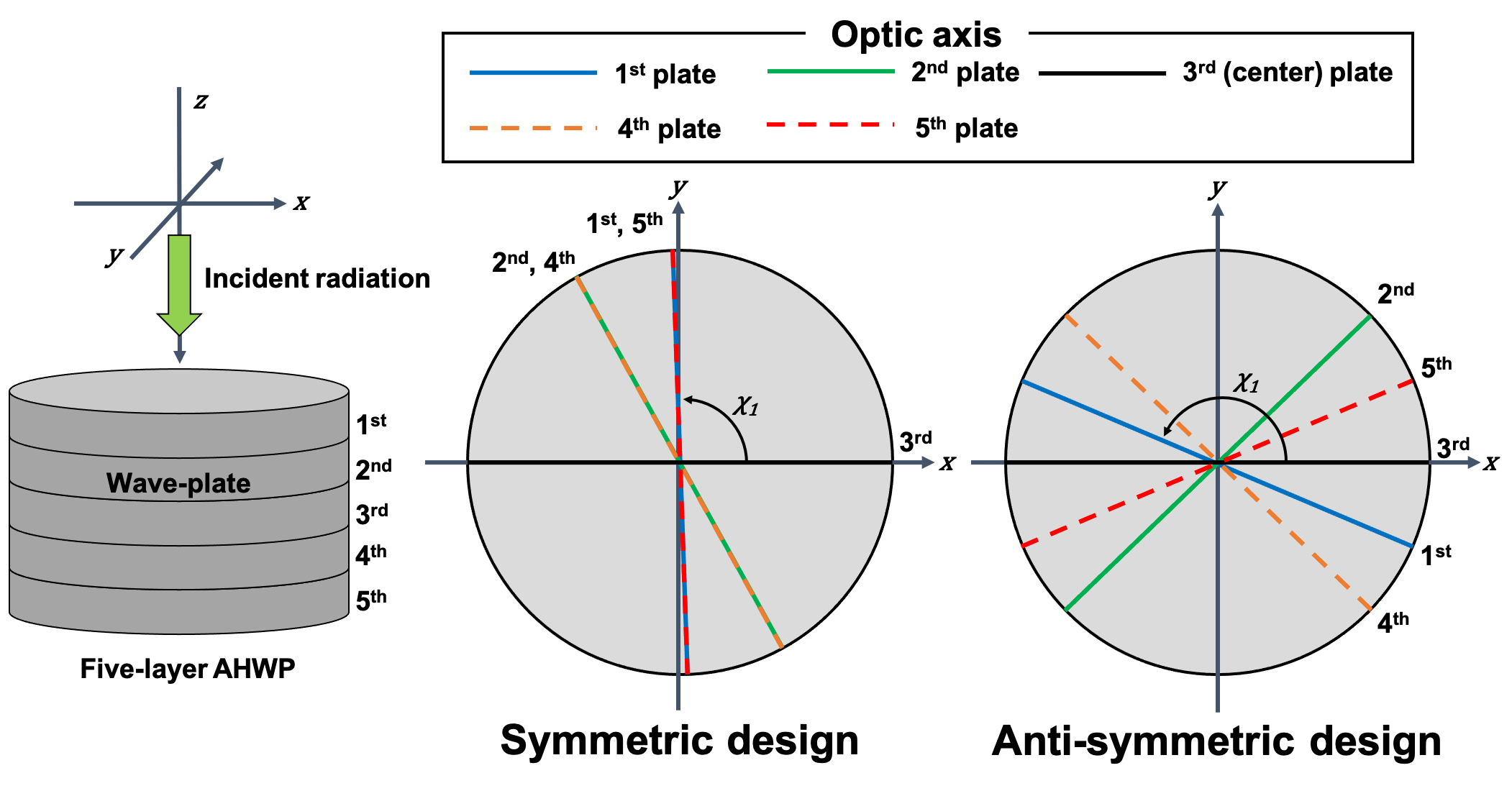}
\end{center}
\caption{Conceptual sketch of the symmetric and anti-symmetric designs for the case of a five layer AHWP.}
\label{fig:concept_asym_sym}
\end{figure}

We present the formalism of HWP polarimetry for our optimization study.
Similar descriptions of the formalism can be found in Komatsu et al.\cite{kkomatsu_jatis_2019}.
To simplify the computational methods, we ignore the effect of reflection on the HWP surface.
The retardance for a single wave plate is defined as 
\begin{eqnarray}
\delta(\nu) = 2\pi \nu \ \frac{d |n_e-n_o|}{c},
\label{eq:retardance}
\end{eqnarray}
where $\nu$ is the frequency, $n_{o}$ and $n_{e}$ are the refractive indices for the ordinary and extraordinary rays, $d$ is the thickness of a single wave plate, and $c$ is the speed of light. 
The Mueller matrix of the wave plate with the retardance of $\delta(\nu)$ and a rotation with an angle of $\chi$~\cite{Shurcliff+2013} are
\begin{equation}
\gamma(\nu)=\left(
\begin{array}{cccc}
1&0&0&0\\
0&1&0&0\\
0&0&\cos\delta(\nu)&-\sin\delta(\nu)\\
0&0&\sin\delta(\nu)&\cos\delta(\nu) 
\end{array}
\right), \ \ \ \ 
R(\chi)=\left(
\begin{array}{cccc}
1&0&0&0\\
0&\cos2\chi&-\sin2\chi&0 \\
0&\sin2\chi&\cos2\chi&0 \\
0&0&0&1 
\end{array}
\right).
\label{mt:rotation}
\end{equation}
The $N$-layered wave plates with a relative angle $\chi_i$ for the $i^{th}$ layer can be written with the rotation matrix $R$ as
\begin{equation}
\Gamma(\nu)=\prod_i^N R(-\chi_{i})\gamma(\nu) R(\chi_{i}).
\label{mt:ahwp_wo_refl}
\end{equation}
The Mueller matrix of $\Gamma$ represents the AHWP. 
When incident radiation with the Stokes parameters, $S_{\rm in}(\nu)=(I_{\rm in}(\nu),Q_{\rm in}(\nu),U_{\rm in}(\nu),V_{\rm in}(\nu))$ in units of spectral radiance, propagates through the continuously-rotating $N$-layered wave plate with an angle $\rho$, the output Stokes parameters, $S_{\rm out}(\nu)=(I_{\rm out}(\nu),Q_{\rm out}(\nu),$ $U_{\rm out}(\nu),V_{\rm out}(\nu))$, is written as
\begin{equation}
S_{\rm out}(\nu)=R(-\rho)\Gamma(\nu) R(\rho)S_{\rm in}(\nu).
\label{eq:Sout}
\end{equation}
Radiation which passes through the AHWP is detected by a linearly polarization-sensitive detector. 
We simplify this step in our modeling by defining a perfect wire grid along the $x$-axis as also shown in Fig.~\ref{fig:concept_asym_sym}, and selecting the intensity component of the final Mueller matrix.
As a result, the detected signal can be written as 
\begin{equation}
    I_{\rm det}(\nu)=(GS_{\rm out}(\nu))  \Bigl|_{I},
\label{eq:Sdet}
\end{equation}
where $G$ is a Mueller matrix of a perfect wire grid along the $x$-axis~\cite{Shurcliff+2013} expressed as,
\begin{equation}
G=\frac{1}{2}\left(
\begin{array}{cccc}
1&1&0&0\\
1&1&0&0 \\
0&0&0&0 \\
0&0&0&0 
\end{array}
\right).
\label{mt:polarizer}
\end{equation}

In this paper, we set $V_{\rm in}(\nu)=0$ by assuming there is no circularly polarized light in the CMB.
The detected intensity $I_{\rm det}(\nu)$ as a function of the HWP angle $\rho$ can be written as 
\begin{equation}
\begin{split}
I_{\rm det}(\nu)=&D_{\rm 0I}(\nu)I_{\rm in}(\nu)+D_{\rm 0Q}(\nu)Q_{\rm in}(\nu)+D_{\rm 0U}(\nu)U_{\rm in}(\nu) \\ 
&+D_{\rm 2I}(\nu)I_{\rm in}(\nu)\cos(2\rho-2\phi_{\rm 0}(\nu))+D_{\rm 2}(\nu)\sqrt{Q_{\rm in}(\nu)^{2}+U_{\rm in}(\nu)^{2}}\cos(2\rho-2\phi_{2}(\nu)) \\
&\qquad \qquad \qquad \qquad \qquad \qquad +D_{\rm 4}(\nu)\sqrt{Q_{\rm in}(\nu)^{2}+U_{\rm in}(\nu)^{2}}\cos(4\rho-4\phi_{\rm 4}(\nu)).
\label{eq:Iout}
\end{split}
\end{equation}
The polarization efficiency and the phase are $2D_{4}(\nu)$ and $\phi_{4}(\nu)$, respectively.
It is worth stressing that these two variables depend on the frequency of the incident radiation for an AHWP while only the polarization efficiency depends on frequency for a single HWP.
Each term in this equation can be related to the Mueller matrix element of $\Gamma(\nu)$ as
\begin{equation}
\begin{split}
D_{\rm 0I}(\nu)&=\frac{1}{2}M_{\rm II}(\nu), \\
D_{\rm 0Q}(\nu)&=\frac{1}{4}(M_{\rm QQ}(\nu)+M_{\rm UU}(\nu)), \\
D_{\rm 0U}(\nu)&=\frac{1}{4}(M_{\rm QU}(\nu)-M_{\rm UQ}(\nu)), \\
D_{\rm 2I}(\nu)&=\frac{1}{2}\sqrt{M_{\rm UI}(\nu)^{2}+M_{\rm QI}(\nu)^{2}}, \\
\phi_{\rm 0}(\nu)&=\frac{1}{2}\arctan\frac{M_{\rm UI}(\nu)}{M_{\rm QI}(\nu)},  \\
D_{\rm 2}(\nu)&=\frac{1}{2}\sqrt{M_{\rm IQ}(\nu)^{2}+M_{\rm IU}(\nu)^{2}}, \\
\phi_{\rm 2}(\nu)&=\frac{1}{2}\arctan\frac{M_{\rm IU}(\nu)}{M_{\rm IQ}(\nu)}+\frac{1}{2}\arctan\frac{U_{\rm in}(\nu)}{Q_{\rm in}(\nu)},  \\
D_{\rm 4}(\nu)&=\frac{1}{4}\sqrt{(M_{\rm QQ}(\nu)-M_{\rm UU}(\nu))^{2}+(M_{\rm QU}(\nu)+M_{\rm UQ}(\nu))^{2}},\\
\phi_{\rm 4}(\nu)&=\frac{1}{4}\arctan\frac{M_{\rm QU}(\nu)+M_{\rm UQ}(\nu)}{M_{\rm QQ}(\nu)-M_{\rm UU}(\nu)}+\frac{1}{4}\arctan\frac{U_{\rm in}(\nu)}{Q_{\rm in}(\nu)}, 
\label{eq:terms}
\end{split}
\end{equation}
where the element of $\Gamma(\nu)$ is generalized as 
\begin{equation}
\Gamma(\nu)=\left(
\begin{array}{cccc}
M_{\rm II}(\nu)&M_{\rm IQ}(\nu)&M_{\rm IU}(\nu)&M_{\rm IV}(\nu) \\
M_{\rm QI}(\nu)&M_{\rm QQ}(\nu)&M_{\rm QU}(\nu)&M_{\rm QV}(\nu) \\
M_{\rm UI}(\nu)&M_{\rm UQ}(\nu)&M_{\rm UU}(\nu)&M_{\rm UV}(\nu) \\
M_{\rm VI}(\nu)&M_{\rm VQ}(\nu)&M_{\rm VU}(\nu)&M_{\rm VV}(\nu) 
\end{array}
\right).
\label{mt:stacked_waveplate}
\end{equation}

\noindent
To account for the finite bandwidth, we integrate the detected intensity over the band as
\begin{equation}
\begin{split}
\left<I_{det}\right>&=\int_{0}^{\infty} w(\nu) I_{det}(\nu) d\nu,
\end{split}
\label{eq:integration1}
\end{equation}
where $w(\nu)$ is the weight function to take into account a detailed band shape.
If we pick up the final term in Eq.~\ref{eq:Iout}, we can write
\begin{equation}
\begin{split}
\int_{0}^{\infty}  w(\nu)D_{\rm 4}(\nu) \sqrt{Q_{\rm in}(\nu)^2+U_{\rm in}(\nu)^2} \cos(4\rho-4\phi_{\rm 4}(\nu)) d\nu
= A_{4}\cos{(4\rho - 4\Phi_{4})}.
\end{split}
\label{eq:integration2}
\end{equation}
The polarization efficiency and phase from $\left<I_{\rm det}\right>$ relate as $2A_4$ and $\Phi_{4}$.
Therefore, we use $2A_{4}$ as the figure of merit and use the relative angles of the wave plates, $\chi_{i}$, as optimization variables. 
The thickness of each plate can be used as another optimization parameter, but we did not find the use in our analysis as described in the later sections. 

\subsection{Optimization}
\label{sec:opt}
In our optimization, we set the condition for the angle $\chi_i$ to be oriented such that each optic axis angle is anti-symmetric with respect to the angle of the center plate as shown in Fig.~\ref{fig:concept_asym_sym} to achieve the frequency-independent phase.
Hereafter, we call this type of AHWP design as ``anti-symmetric design".
This condition is key to realize the frequency-independent optic axis over a broad band.
We use a five layer anti-symmetric design as an example to describe how the polarization angle rotates through an AHWP. 
In the anti-symmetric design, the first and fifth plates are rotated by the same angle in opposite directions about the orientation of the third axis. The fourth and second plates are rotated similarly, but at a different angle than the first and fifth plates.
Namely, we impose the conditions of $\chi_4=-\chi_2$ and $\chi_5=-\chi_1$ in case of $N=5$. 
In this way, any offset rotational angle introduced by the first 2.5 plates at each frequency is canceled by the second 2.5 plates.
While we achieve the cancellation of the frequency-dependent offset angle we still achieve the rotation of the incoming linear polarization plane as the AHWP rotates. 
A similar concept can be found for the case which two sets of the AHWPs are employed to cancel the frequency dependent effective fast axis response.~\cite{roc_ahwp}.
With this condition, we do not randomize all the $\chi_i$ angles fully, but randomize only the $(N-1)/2$ plates of the $N$-layer AHWP.

In the following, we demonstrate the optimization of the AHWP design by using this anti-symmetric constraint with a specific observational frequency range. 
LiteBIRD~\cite{hazumi_spie2020} is an example project to apply our AHWP designs.
The LiteBIRD low frequency telescope~\cite{sekimoto_spie2020} has the largest frequency coverage, 34-161~GHz, within a single telescope with a presence of a HWP to date. Thus, we use this range in our study as an example.
We assume a top-hat band shape in this frequency range. 
In a real application, we should be able to derive the optimized recipe by taking into account a more detailed band shape, but this is beyond the scope of this paper. 
In this work, we use sapphire as a wave plate. 
The set of the parameters used in this design work is summarized in Tab.~\ref{tab:fix_param}.
 \begin{table}[bt]
\caption{\label{tab:fix_param} A summary of the parameters used for the optimization process.} 
\begin{center}   
\begin{tabular}{c|c} 
    	 Frequency range, $\nu$ & $34-161$~GHz \\ \hline 
	     Center frequency, $\nu_{0}$ & 97.5~GHz \\ \hline 
    	 Refractive indices, $(n_{o}, n_{e})$  & $(3.047, 3.361)$~\cite{b.r.johnson} \\ \hline 
         Thickness of each plate, $d$  & 4.9~mm \\ \hline
         Incident angle   & 0~degrees \\ 
\end{tabular}
\end{center}
\end{table}

In the optimisation calculation, the integration of Eq.~\ref{eq:integration2} is replaced by summation of Eq.~\ref{eq:Iout} with the frequency resolution of 1~GHz.
We calculate $A_{4}$ a total of 50,000 times with random distribution of $\chi_{i}$ between 0 and 180 degrees for $S_{in}=(1,0,1,0)$, and find the optimal designs for 3, 5, 7, 9 layers.
The final result does not depend on a detailed choice of $Q_{\rm in}$ and $U_{\rm in}$ except for an offset of $\phi_4$.
From this calculation, we choose a set of $\chi_{i}$ angles that provide the largest $A_{4}$ over the given frequency range.

We have only explored the case with an odd number of wave plates because of the existing broadband AHWP designs from past studies~\cite{kkomatsu_jatis_2019,hanany,savini,pisano}. 
It is worth pointing out that these existing designs tend to have relative angles oriented symmetrically with respect to the angle of the center plate. 
In order to compare the results of our anti-symmetric design to existing designs, we also compute the case in which the optic axis angles are set to be symmetric with respect to the angle of the center plate as Fig.~\ref{fig:concept_asym_sym}. 
Hereafter, we call this type of AHWP design as the ``symmetric design".

\section{Results}
\label{sec:results}
\begin{figure}[htb]
\begin{center}
\includegraphics[width=\hsize]{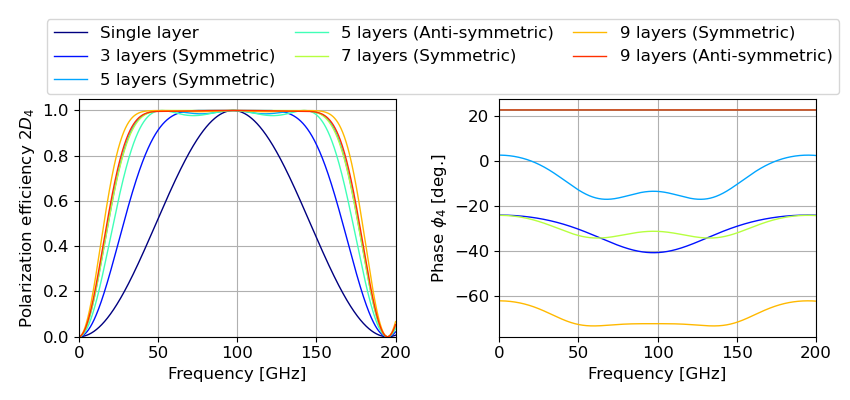}
\end{center}
\caption{The comparison of polarization efficiency and phase of each design. 
In the phase plot, the line of the 5 layer anti-symmetric case is under the line of the 9 layer anti-symmetric case.}
\label{fig:compare_designs}
\end{figure}

Tab.~\ref{tab:asym} and \ref{tab:sym} show the results of the parameter searches. 
We list the fractional bandwidth $\Delta\nu/\nu_{0}$ that is the ratio of the frequency range with polarization efficiency $2D_{4}$ greater than 0.9 to a center frequency of 97.5~GHz.
We also list the band-averaged polarization efficiency $2A_{4}$ and the maximum phase differences $\Delta\phi_{4}$ are computed within the frequency range in Tab.~\ref{tab:fix_param}.
Fig.~\ref{fig:compare_designs} shows polarization efficiency and phase $\phi_{4}$ as a function of the frequency based on the designs listed in Tab.~\ref{tab:asym} and \ref{tab:sym} for $S_{in}=(1,0,1,0)$.
The symmetric angle case shows a broad coverage of the polarization efficiency and the non-flat phase response over frequency. 
On the other hand, the anti-symmetric design for $N=5$ and $9$ can completely eliminate the phase variation over frequency while the polarization efficiency is maintained to be broad. 
Demonstrations of the $N=5$ anti-symmetric design assembled with an optic axis alignment accuracy of 1 degree can be found in Komatsu et al.~\cite{kkomatsu_spie_demo_2020}

The quoted band width of polarization efficiency in Tab.~\ref{tab:asym}, and \ref{tab:sym} should be treated as a representative value. 
Depending on the application, one might allow to have some oscillatory features around the polarization efficiency close to 1 as the $N=5$ anti-symmetric design in Fig~\ref{fig:compare_designs}.
In this case, there is a possibility that we obtain a broader bandwidth by trading the degradation of the overall averaged efficiency. 
In such a case, we can increase the band width. 
This point is addressed in Sec.~\ref{sec:opt_wider_freq}. 

We also computed for three and seven anti-symmetric angles. 
For $N=3$, we found the solution to be $\chi_{i} = (90.00, 0.00, -90.00)$ degrees, which is essentially the same as a single HWP. 
For $N=7$, we found the solution to be $\chi_{i} = (111.73, 43.70, 97.20, 0.00, -97.20, -43.70, -111.73)$ degrees, but the calculated polarization efficiency and fractional bandwidth are nearly the same as $N=5$. 
Therefore, we omit to show these in the table and figure due to redundancy.

The solutions which we have shown in this paper are not unique.
For example, with the angle set for the $N=5$ anti-symmetric design, the same performance can be obtained by the angle set of
\begin{eqnarray}
    \chi_1 &=& 22.67 \pm 180\times j\\
    \chi_2 &=& 133.63 \pm 180\times j\\
    \chi_3 &=& 0.00 \pm 180\times j\\
    \chi_4 &=& -133.63 \pm 180\times j\\
    \chi_5 &=& -22.67 \pm 180\times j
\end{eqnarray}
in units of degrees, where $j$ is an arbitrary integer because of the spin-2 nature of the wave plate.
Therefore, the set of angles may look different, but multiple combinations of angles can produce the same performance. 
Needless to say, there are overall rotational degrees of freedom, thus any global rotation, i.e. $\chi_3\neq0$, added to all the angles $\chi_i$ still provides the same spectral performance except for the change of the global phase offset $\phi_4$.

\begin{table}[h]
\caption{\label{tab:asym} A summary of anti-symmetric designs. The the maximum phase difference $\Delta\phi_{4}$ is completely zero.} 
\begin{center}   
\begin{tabular}{c|c|c|c|c} 
    	The number &  Fractional & Polarization  & Phase & Optic axis   \\
        of layers & bandwidth & efficiency & difference &  angles  \\
    	$N$ & $\Delta\nu/\nu_{0}$ & $2A_{4}$ & $\Delta\phi_{4}$[deg.] & $\chi_{i}$ [deg.] \\ \hline\hline 
    	   5  & 1.23 & 0.978 & 0.0 & 22.67 , 133.63 , 0.00 , -133.63 , -22.67  \\  \hline
    	   9 & 1.35 & 0.993 & 0.0 & \begin{tabular}{c} 23.19 , 170.88 , 89.85 , 143.85 , 0.00 , \\ -143.85 , -89.85 , -170.88 , -23.19  \end{tabular} \\ 
\end{tabular}
\end{center}
\end{table}

\begin{table}[h]
\caption{\label{tab:sym} Table of symmetric designs} 
\begin{center}   
\begin{tabular}{c|c|c|c|c} 
    	The number &  Fractional & Polarization  & Phase & Optic axis   \\
        of layers & bandwidth & efficiency & difference &  angles  \\
    	$N$ & $\Delta\nu/\nu_{0}$ & $2A_{4}$ & $\Delta\phi_{4}$[deg.] & $\chi_{i}$ [deg.] \\ \hline\hline 
    	   3  & 1.00 & 0.894 & 13.9 & 58.35 , 0.00 , 58.35   \\   \hline 
    	   5  & 1.33 & 0.965 & 11.9 & 88.65 , 61.68 , 0.00 , 61.68 , 88.65    \\  \hline
    	   7 & 1.32 & 0.989 & 5.1 & \begin{tabular}{c} 49.76 , 99.86 , 23.20 , 0.00 , 23.20 , \\ 99.86 , 49.76  \end{tabular} \\ \hline
    	   9 & 1.42 & 0.993 & 4.8 & \begin{tabular}{c} 1.83 , 66.92 , 15.36 , 132.66 , 0.00 , \\ 132.66 , 15.36 , 66.92 , 1.83  \end{tabular} \\ 
\end{tabular}
\end{center}
\end{table}

\section{Discussions}

\subsection{Tolerance analysis}
LiteBIRD plans to use a five layer AHWP instead of a nine layer AHWP due to the requirement for HWP weight and so on~\cite{ysakurai_spie2020}.
Therefore, as an example, we discuss alignment tolerances for the $N=5$ anti-symmetric design.

\begin{figure}[b]
\begin{center}
\includegraphics[width=\hsize]{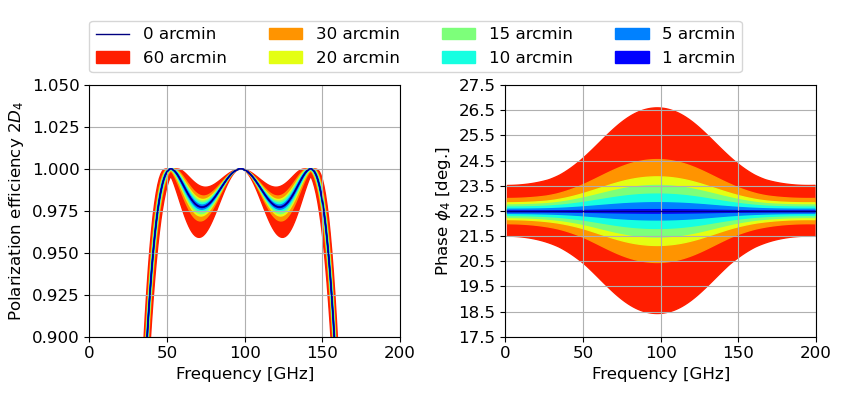}
\end{center}
\caption{The relation between alignment accuracy of the optic axis angles of each plate and AHWP performance of the $N=5$ anti-symmetric design. The left panel shows polarization efficiency and the right panel shows phase. }
\label{fig:error_prop_angle}
\end{figure}

\subsubsection{Optic axis angle alignment}
Fig.~\ref{fig:error_prop_angle} shows the relation between the alignment accuracy of $\chi_i$ and AHWP performances, polarization efficiency, and phase, in the case of the $N=5$ anti-symmetric design.
We fix all the parameters except for the relative HWP angles $\chi_i$, which are randomized with flat distributions for each plate in the range of 1, 5, 10, 15, 20, 30 and 60~arcmin centered at the nominal designed angles.
For each case, we compute polarization efficiency and phase.
We perform this calculation 10,000 times in each case to obtain the range of the performances. 
Fig.~\ref{fig:error_prop_angle} shows the susceptibility to the frequency dependence of the effective fast axis, phase $\phi_4$, from the alignment accuracy of each optic axis angle.
To suppress the maximum deviation of the frequency dependent phase to be less than 1~degree, the alignment accuracy of the optic axis angle is required to be less than 15~arcmin.

The alignment accuracy of the rotational angle between the plates can be achieved to $<10$~arcmin~\cite{kkomatsu_jatis_2019} by aligning the orientation flat between the plates using a properly designed alignment jig, e.g. a universal measurement machine. 
The orientation flat can be machined to each plate, and its accuracy can be about a degree without any effort and can be sub-degree level if the crystal orientation is determined by using X-ray diffraction. 

\subsubsection{Wave plate thickness}
\begin{figure}[b]
\begin{center}
\includegraphics[width=\hsize]{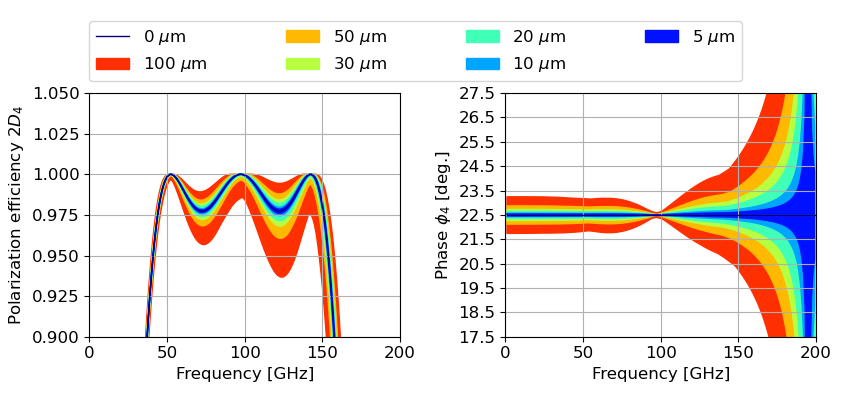}
\end{center}
\caption{The relation between the variations of wave plate thickness and AHWP performances in the N=5 anti-symmetric design five layer AHWP.}
\label{fig:error_prop_thickness}
\end{figure}

We also perform the tolerance analysis to the accuracy of the wave plate thickness.
Fig.~\ref{fig:error_prop_thickness} shows the polarization efficiency and the phase when we add the flat distribution of a random thickness within the range of $5, 10, 20, 30, 50,$ and $100~\mu$m to the nominal thickness for the $N=5$ anti-symmetric design.
Each plate thickness is varied independently without any correlation among the plates.
We assume that the plates are always adjacent to each other without any gap. 
From these figures, we identify that the impact is prominent at the higher side of the frequency range.
The reason is simply because the higher frequency is more susceptible to small changes of the thickness in order to maintain the same retardance as shown in Eq.~\ref{eq:retardance}.
A similar impact appears in the phase. 
At 195~GHz, the retardance with the nominal thickness is $2\pi$, and thus $A_4=0$. 
Correspondingly, no phase can be defined. 
The impact of the uncertainty to the phase is also propagated from the last equation in Eq.~\ref{eq:terms} because the element of the Mueller matrix $\Gamma$ is a function of the retardance.
In order to minimize the maximum phase variation in the frequency range from 34 to 161~GHz to be less than 1~degree, thickness variation has to be controlled to be less than 20~$\mu$m.

Sapphire disks with a diameter of 50~mm or larger are commercially available with an accuracy of 0.1~mm. 
The measurement accuracy can be higher than 10~$\mu m$ without serious effort, and therefore the AHWP designer can account for this thickness variation as a part of the input design. 
The surface accuracy of a sapphire disk is nominallu a few tens of $\mu m$ for 50~mm size diameter, but this can be challenging to maintain small as the diameter becomes larger, e.g. $300~$mm. 
Toda et al.~\cite{toda_spie2020} shows the 7~$\mu m$ surface accuracy for a sapphire disk plate with a diameter of 50~mm. 
Komatsu et al.~\cite{kkomatsu_jatis_2019} found the surface accuracy of sapphire with a diameter of 100~mm to be less than 8~$\mu m$.

\begin{table}[b]
\caption{\label{tab:asym2_5} Table of $N=5$ anti-symmetric designs optimized with various frequency ranges. The polarization efficiencies for the top three frequencies deviate from 1 by orders of $10^{-6}$, $10^{-5}$, and $10^{-4}$, respectively. We omit the phase difference since phase is frequency-independent in the anti-symmetric design.}
\begin{center}   
\begin{tabular}{c|c|c|c|c} 
     	Optimization & Optimization & Fractional & Polarization & Optic axis   \\ 
     	freq. range & bandwidth & bandwidth & efficiency & angles \\ 
    	$\Delta\nu_{opt}$  [GHz] & $\Delta\nu_{opt}/\nu_{0}$ & $\Delta\nu/\nu_{0}$ & $2A_{4}$ &  $\chi_{i}$ [deg.] \\ \hline\hline 
        $ 84 - 111 $ & 0.28 & 1.02 &1.000 & 23.28 , 128.13 , 0.00 , -128.13 , -23.28   \\   \hline
        $ 74 - 121 $ & 0.48 & 1.04 &1.000 & 157.34 , 51.92 , 0.00 , -51.92 , -157.34   \\   \hline
        $ 64 - 131 $ & 0.69 & 1.09 &1.000 & 23.11 , 129.59 , 0.00 , -129.59 , -23.11   \\   \hline
        $ 54 - 141 $ & 0.89 & 1.12 &0.998 & 23.55 , 130.70 , 0.00 , -130.70 , -23.55   \\   \hline
        $ 44 - 151 $ & 1.10 & 1.17 &0.993 & 156.95 , 48.05 , 0.00 , -48.05 , -156.95   \\   \hline
        $ 34 - 161 $ & 1.30 & 1.23 &0.978 & 22.67 , 133.63 , 0.00 , -133.63 , -22.67   \\   \hline
        $ 24 - 171 $ & 1.51 &1.28 &0.945 &  157.83 , 45.21 , 0.00 , -45.21 , -157.83   \\   \hline
        $ 14 - 181 $ & 1.71 & 1.31 &0.886 & 157.98 , 44.07 , 0.00 , -44.07 , -157.98   \\   \hline
        $ 4 - 191 $ & 1.92 & 1.33 &0.806 & 21.51 , 136.51 , 0.00 , -136.51 , -21.51   \\  
\end{tabular}
\end{center}
\end{table}

\begin{table}[b]
\caption{\label{tab:asym2_9} Table of $N=9$ anti-symmetric designs optimized with various frequency ranges. The polarization efficiencies for the top three frequencies deviate from 1 by orders of $10^{-6}$, $10^{-5}$, and $10^{-4}$, respectively. We omit the phase difference since phase is frequency-independent for the anti-symmetric design.} 
\begin{center}   
\begin{tabular}{c|c|c|c|c} 
     	Optimization & Optimization & Fractional & Polarization & Optic axis   \\ 
     	freq. range & bandwidth & bandwidth & efficiency & angles \\ 
    	$\nu_{opt}$ [GHz] & $\Delta\nu_{opt}/\nu_{0}$ & $\Delta\nu/\nu_{0}$ & $2A_{4}$ &  $\chi_{i}$ [deg.] \\ \hline\hline 
        $84 - 111$ & 0.28 & 0.96 & 1.000 & \begin{tabular}{c} 11.71 , 154.96 , 56.67 , 66.41 , 0.00 , \\ -66.41 , -56.67 , -154.96 , -11.71  \end{tabular}  \\   \hline
        $74 - 121$ & 0.48 & 1.07 & 1.000 & \begin{tabular}{c} 50.10 , 142.71 , 19.65 , 124.47 , 0.00 , \\ -124.47 , -19.65 , -142.71 , -50.10  \end{tabular}  \\   \hline
        $64 - 131$ & 0.69 & 1.08 & 1.000 & \begin{tabular}{c} 145.87 , 52.56 , 17.93 , 122.93 , 0.00 , \\ -122.93 , -17.93 , -52.56 , -145.87  \end{tabular}  \\   \hline
        $54 - 141$  & 0.89 & 1.21 & 0.999 & \begin{tabular}{c} 0.73 , 139.90 , 42.67 , 60.88 , 0.00 , \\ -60.88 , -42.67 , -139.90 , -0.73  \end{tabular}  \\   \hline
        $44 - 151$  & 1.10 & 1.35 & 0.998 & 23.19 , 170.88 , 89.85 , 143.85 , 0.00,     \\   \cline{1-2} \cline{4-4}
        $34 - 161$ & 1.30 &  & 0.993 & -143.85 , -89.85 , -170.88 , -23.19 \\   \hline
        $24 - 171$  & 1.51 & 1.50 & 0.984 & \begin{tabular}{c} 158.36 , 166.52 , 65.73 , 35.57 , 0.00 , \\ -35.57 , -65.73 , -166.52 , -158.36  \end{tabular}  \\   \hline
        $14 - 181$ & 1.71 & 1.55 & 0.957 &  20.64 , 4.72 , 108.96 , 150.26 , 0.00 ,    \\   \cline{1-2} \cline{4-4}
        $4 - 191$ & 1.92 &  & 0.886 &  -150.26 , -108.96 , -4.72 , -20.64  \\   
\end{tabular}
\end{center}
\end{table}

\begin{figure}[h]
\begin{center}
\includegraphics[width=\hsize]{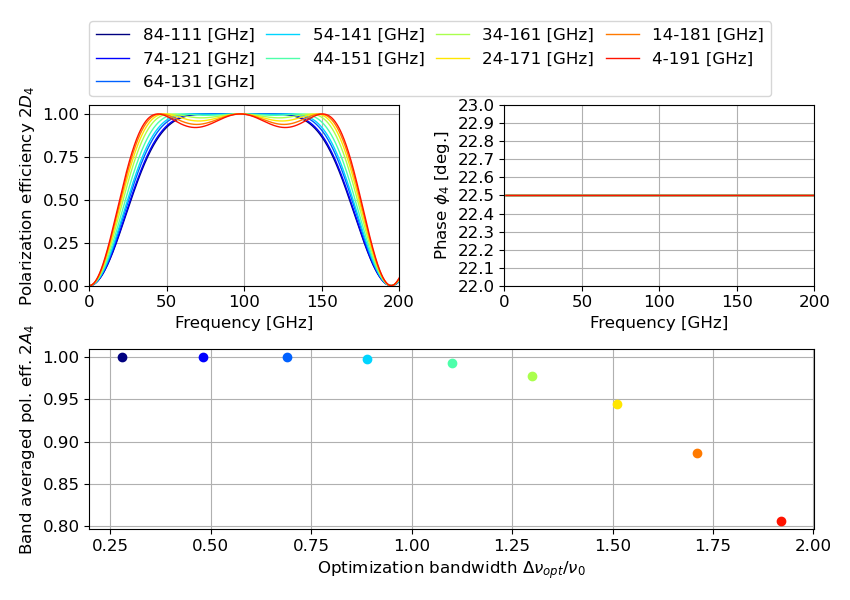}
\end{center}
\caption{The comparison of the polarization efficiency and the effective fast axis angle of Tab.~\ref{tab:asym2_5}. 
The horizontal axis of the bottom plot is the ratio of the optimization frequency range to the canter frequency.
In the phase plot, all lines overlap since the all designs have the same value.}
\label{fig:compare_asym5_designs_wider}
\end{figure}

\subsection{Further optimization for broader frequency coverage}
\label{sec:opt_wider_freq}
When we allow an oscillatory feature near a polarization efficiency of 1, we can broaden the bandwidth as described in Sec.~\ref{sec:results}. 
Conversely, when we do not allow an oscillatory feature near a polarization efficiency of 1, the available bandwidth becomes narrower. 
As a test case, we set the varieties of the frequency range to be optimized. 
Specifically, we start with a frequency range of 34-161 GHz and add/subtract 10 GHz to widen/narrow the band width.
We optimise with nine frequency ranges for the $N=5$ and $9$ anti-symmetric design.
Tab.~\ref{tab:asym2_5} and ~\ref{tab:asym2_9} show the results of the optimization for each frequency range.
We define the optimization fractional bandwidth as $\Delta\nu_{opt}/\nu_{0}$, which is the ratio of the targeted optimization range $\Delta\nu_{opt}$ to the center frequency of 97.5~GHz.
Fig.~\ref{fig:compare_asym5_designs_wider} shows the polarization efficiency, phase, and band averaged polarization efficiency over the optimization bandwidth for all the cases of $N=5$.
From the top-left panel of Fig.~\ref{fig:compare_asym5_designs_wider}, we confirm that the optimized design has larger oscillatory features at polarization efficiency close to 1 when we use broader bandwidth to be optimized. 
An AHWP designer has to take into account the trade-off between the broadband availability and the overall averaged polarization efficiency based on each application.
This consideration applies the same for both the symmetric and anti-symmetric designs. 

\subsection{Further design optimization with larger degrees of freedom}
\begin{figure}[htb]
\begin{center}
\includegraphics[width=\hsize]{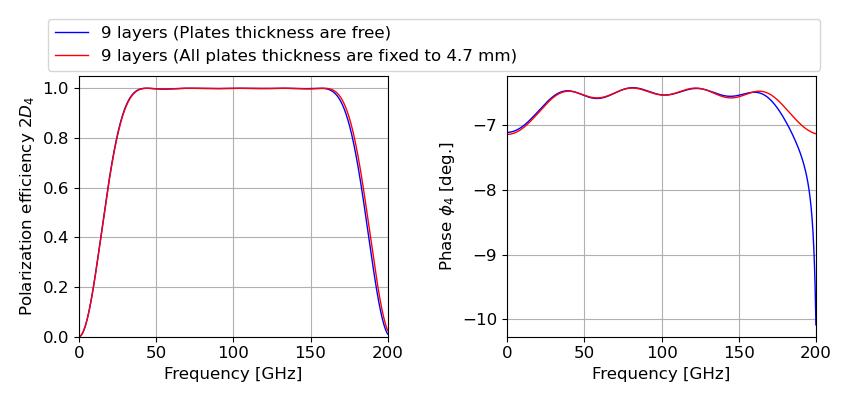}
\end{center}
\caption{Polarization efficiency and effective fast axis angle of the design in Tab.~\ref{tab:nk}.}
\label{fig:compare_nk}
\end{figure}

The optimizations were carried about by assuming a fixed thickness of each wave plate and by imposing the anti-symmetric condition.
Here, we do not enforce these conditions. 
As a result, we take the thickness of all wave plates and all relative angles as free parameters and carry out optimization with a figure of merit of $2A_4$ for the frequency range of 34-161~GHz.
We choose $N=9$, which gives $8+9$ extra free parameters. 
Because of this large number of free parameters, we use minuit~\cite{imimuit} to maximize $A_{4}$ after randomizing the parameters for the optimization in this section.
Tab.~\ref{tab:nk} shows the results of the optimization. 
All thicknesses ended up converging to essentially the same value. 
Fig.~\ref{fig:compare_nk} shows polarization efficiency and phase as a function of frequency with the parameters in Tab.~\ref{tab:nk} and the case with thickness being fixed at 4.7~mm.
While we had assumed that adding free parameters should increase the degrees of freedom to find a broader polarization efficiency with a flat phase response, we did not find the broader coverage within the range of our parameter searches. 
Due to the large number of the free parameters, there may be room for improvement in the optimization process, but such an investigation is beyond the scope of this paper. 
However, one of the designs shown in Tab.~\ref{tab:nk} has a maximum phase difference of about 0.2 degrees over the targeted frequency range without imposing an anti-symmetric condition. 
This means an anti-symmetric condition is not the only way to reach a flat phase response when we have a large number of free parameters.

\begin{table}[b]
\caption{\label{tab:nk} Table of an alternative design with a frequency-dependent optic axis} 
\begin{center}   
\begin{tabular}{c|c|c|c|c|c} 
    	Number  & Fractional & Polarization & Phase & \multicolumn{2}{c}{}  \\ 
        of layers & bandwidth & efficiency & difference & \multicolumn{2}{c}{}  \\ 
    	$N$ & $\Delta\nu/\nu_{0}$ & $2A_{4}$ & $\Delta\phi_{4}$[deg.] & \multicolumn{2}{c}{} \\ \hline\hline 
   	  9 & 1.48 & 0.998 & 0.17 & $\chi_{i}$ &  -69.92,   5.13,  -7.27,  44.90 ,   0.00,  \\ 
       &  & & & [deg.] &  109.89, -18.27, -36.60 ,
        29.83 \\ \cline{5-6}
       & & & & $d_{i}$ & 4.694, 4.649, 4.755, 4.686, 4.747, \\ 
       & & & & [mm] & 4.700, 4.743, 4.663, 4.713
\end{tabular}
\end{center}
\end{table}

\section{Conclusions}
\label{sec.conclusion}
A HWP for CMB polarization experiments needs to cover a broad band in order to achieve high sensitivity to CMB radiation as well as foreground emissions in a single polarimeter. 
While an AHWP can broaden the band width, the frequency-dependent fast axis of a AHWP can add challenges to calibration and analysis complexity. 
We propose eliminating this effect with a specific angle set of a novel AHWP by imposing the anti-symmetric orientation to the relative wave plate angles. 
We derived the examples of the wave plate relative angles for $N=5$ and $N=9$. 
The optimized set of relative angles achieves the frequency-independent optic axis and covers a fractional bandwidth of 1.3 and 1.5 for five and nine layer AHWPs, respectively.
We also discussed the tolerance of the design in relation to wave plate relative angles and thicknesses. 
In order to minimize the maximum variation of phase response over frequency, we need to assemble the AHWP within relative angles of 15~arcmin and with thickness accuracy less than $20~\mu$m.
This result can be applicable not only to CMB polarimetry, but any other application which requires flat spectral response of the effective fast axis of an AHWP. 

\section*{Acknowledgments}
This paper based on a SPIE conference proceedings paper~\cite{kkomatsu_spie_design_2020}.
This work was supported 
by JPSP KAKENHI Grant Number JP17H01125, JP18J20148, JP19K14732, by World Premier International Research Center Initiative (WPI), MEXT,
and by the JSPS Core-to-Core program, A. Advanced Research Networks. 
We would like to thank Dr. Samantha Stever for editorial suggestions to this paper.



\bibliography{report}   
\bibliographystyle{spiejour}   

\vspace{2ex}\noindent\textbf{First Author} is a PhD student at Okayama University. He received his BS and MS degrees in physics from Okayama University in 2016 and 2018, respectively. His current research interests the verification of inflation theory using B-mode polarization of CMB created by the primordial gravitational waves. Related to it, he is developing the polarization modulator of LiteBIRD.


\listoffigures
\listoftables

\end{spacing}
\end{document}